\documentclass[11pt]{article}
\usepackage{amsmath,amssymb,amsfonts}
\usepackage[english]{babel}
\usepackage{float}
\makeatletter
\@addtoreset{equation}{section}
\makeatother

\def\nn{\nonumber} \def\bd{\begin{document}} \def\ed{\end{document}}
\def\ds{\documentstyle}
\let\bm=\bibitem
\newcommand{\be}{\begin{equation}}
\newcommand{\ee}{\end{equation}}
\newcommand{\bea}{\setlength\arraycolsep{2pt} \begin{eqnarray}}
\newcommand{\eea}{\end{eqnarray}}
\newcommand{\hoch}[1]{$\, ^{#1}$}
\def\p{\partial}
\title{\large {\bf Constructing $p,n$-forms from $p$-forms via the Hodge star operator
and the exterior derivative}}
\date{}
\author{Jun-Jin Peng$^{1,2}$\footnote{pengjjph@163.com}  \\ \\
\small \sl $^1$School of Physics and Electronic Science, Guizhou Normal University,\\
\small Guiyang, Guizhou 550001, People's Republic of China; \\
\small \sl  $^2$Guizhou Provincial Key Laboratory of Radio Astronomy and
Data Processing, \\
\small \sl Guizhou Normal University, \\
\small Guiyang, Guizhou 550001, People's Republic of China
}

\begin{document}

\maketitle
\vspace{20pt}

\begin{center}
\textbf{Abstract}
\end{center}
In this paper, we aim to explore the properties and applications on the operators consisting
of the Hodge star operator together with the exterior derivative, whose action on an
arbitrary $p$-form field in $n$-dimensional spacetimes makes its form degree remain
invariant. Such operations are able to generate a variety of $p$-forms
with the even-order derivatives of the $p$-form. To do this, we first investigate the
properties of the operators, such as the Laplace-de Rham operator,
the codifferential and their combinations, as well as the applications of the
operators in the construction of conserved currents. On basis of two general
$p$-forms, then we construct a general
$n$-form with higher-order derivatives. Finally, we propose that such an $n$-form
could be applied to define a generalized Lagrangian with respect to a $p$-form field
according to the fact that it incudes the ordinary Lagrangians for the $p$-form and
scalar fields as special cases.

\emph{Keywords:} $p$-form; Hodge star; Laplace-de Rham operator;
Lagrangian for $p$-form.

%Mathematics Subject Classification 2010: 53A45, 53B50, 83C50

%%%%%%%%%%%%%%%%%%%%%%%%%%%%%%%%%%%%%%%%%%%%%%%%%%%%%%%%%%%%%%%%%%%%%%%%%
\voffset=-.90pt
\vspace{40pt}

%%%%%%%%%%%%%%%%%%%%%%%%%%%%%%%%%%
\section{Introduction}\label{one}
%%%%%%%%%%%%%%%%%%%%%%%%%%%%%%%%%%

Differential forms are a powerful tool developed to deal with the calculus in
differential geometry and tensor analysis. Their applications in mathematics
and physics have brought about the increasingly widespread attention
\cite{TFrankel,GRStrauman,Waldgr,EGHPhyR}. Particularly, in multifarious
branches of theoretical physics, such as general relativity, supergravity,
(super)string theories, M-theory and so on, antisymmetric tensor fields
are essential ingredients of these theories, while each of such fields is
naturally in correspondence to a certain $p$-form (with the form degree
$p=0,1,2,\cdot\cdot\cdot$). As a consequence, the introduction of
$p$-forms is able to offer great conveniences for the manipulation of
antisymmetric tensor fields, even of various quantities that just contain
antisymmetric parts. For instance, in some cases, if the Lagrangian of an
$n$-dimensional theory is put into an $n$-form, the analysis on the
symmetry of the theory becomes more convenient.

As is known, apart from the wedge product, both the Hodge star operator and the
exterior derivative are thought of as two important and fundamental
tools for manipulating differential forms. Their combinations can generate
various useful non-zero operations \cite{PengHg,JFPleban,SharEg,SharEg2}, such as the
well-known Laplace-de Rham operator, codifferential (divergence operator)
and d'Alembertian operation.
Furthermore, if letting an arbitrary combined operation act upon any $p$-form
field in $n$-dimensional (pseudo-)Riemannian manifold, one observes that the operator
is able to generate $l$-forms only with six types of form degrees, that is, $l\in\{p,q,p\pm1,q\pm1\}$,
where and in what follows $q=n-p$ \cite{PengHg}. However, in the present work, of particular interest
are operators that mix the Hodge star operation with the exterior derivative
and can generate $p$-forms from a $p$-form because of the significance and relevance of these fields
to general relativity and gauge theory. To move on, it is of great necessity
to find all the probable structures of such operators first, as well as to
exploit their main properties then. It will be demonstrated below that the
degree-preserving operations can be expressed through the Laplace-de Rham
operator together with the combinations of the codifferential and the
exterior derivative.

As a matter of fact, working with the action of the combined operators
preserving the form degree upon a certain $p$-form field, one can observe that
all the newly-generated $p$-forms are just the even-order derivatives of this
field. Furthermore, these higher-order derivative $p$-forms, as well as their
Hodge dualities, enable us to construct $n$-forms involving higher derivatives of fields. As a
significant application, in terms of the fact that the ordinary Lagrangians
for the $p$-form gauge fields and scalars can be included as special cases
of those $n$-forms, they may be adopted as appropriate candidates for
Lagrangians with respect to $p$-form
fields. If so, such a proposition would provide a novel understanding towards the $p$-form
gauge theories. For the sake of clarifying this, we recognize that at least the
structure of the $n$-forms, together with their main characters, should be
illustrated at the mathematical level. This is just our main motivation.

The remaining part of the current paper is organized as follows. In section
\ref{two}, in terms of the action upon arbitrary $p$-form fields, we plan to
carry out detailed investigations on the operators generating $p$-forms. We
shall pay special attention to the Laplace-de Rham operator, together with
the codifferential. In section \ref{three}, a general $n$-form, as well
as its equivalent, will be constructed on basis of two general $p$-forms. In
addition, their properties will be analyzed in detail. In section \ref{four},
inspired with the forms of the ordinary Lagrangians for $p$-form fields and
scalars, we are going to put forward that the $n$-forms could play the role of
the generalized Lagrangians with respect to $p$-forms. Then the equations of
motion for the fields will be derived. The last section is our conclusions.
For convenience to the reader, we summarize our notations and conventions
in appendix \ref{appendA}.

%%%%%%%%%%%%%%%%%%%%%%%%%%%%%%%%%%%%%%%%%%%%%%%%%%%%%%%%%%%%%%%%%%%%%%%%
\section{The operators preserving the form degree}\label{two}
%%%%%%%%%%%%%%%%%%%%%%%%%%%%%%%%%%%%%%%%%%%%%%%%%%%%%%%%%%%%%%%%%%%%%%%%

In this section,  through the combination of the Hodge star operation $\star$ and the
exterior derivative ${\rm d}$, we shall present a general operator that can generate differential
forms of degree $p$ from an arbitrary $p$-form field $\textbf{F}$, which is expressed as
$\textbf{F}=(p!)^{-1}F_{\mu_1\cdot\cdot\cdot\mu_p}{\rm d}x^{\mu_1\cdot\cdot\cdot\mu_p}$ (From here
on, ${\rm d}x^{\mu_1\cdot\cdot\cdot\mu_p}$ will always refer to the abbreviation for
${\rm d}x^{\mu_1}\wedge\cdot\cdot\cdot\wedge {\rm d}x^{\mu_p}$ and
$F_{\mu_1\cdot\cdot\cdot\mu_p}$ denotes a totally anti-symmetric rank-$p$
tensor). Further to illustrate such an operator, the Laplace-de Rham operator
$\Delta$, which could be expressed by means of the d'Alembertian (wave)
operator $\Box$ as well as the codifferential $\hat{\delta}$, will be
discussed in detail.

Without loss of generality, in an $n$-dimensional spacetime, which is a pseudo-Riemannian
manifold $M$ endowed with the metric $g_{\mu\nu}$ of a Lorentzian signature
$(-,+,\cdot\cdot\cdot,+)$, we introduce Hodge star that is
also referred to as the Hodge duality operator as the map from $p$-forms to $q$-forms
($q=n-p$), that is, \cite{TFrankel,GRStrauman,Waldgr}
\be
\star :\Omega^p(M)\rightarrow \Omega^q(M)
\, , \label{DefHodge1}
\ee
where $\Omega^p(M)$ stands for the space of $p$-forms on $M$. More specifically,
by means of its action on the $p$-form $\textbf{F}\in \Omega^p(M)$, we obtain
the following $q$-form
\be
\star \textbf{F}= \frac{1}{p!q!}F^{\nu_1\cdot\cdot\cdot\nu_p}
\epsilon_{\nu_1\cdot\cdot\cdot\nu_p\mu_1\cdot\cdot\cdot\mu_q}
{\rm d}x^{\mu_1\cdot\cdot\cdot\mu_q}
\, . \label{DefHodge}
\ee
Here the completely anti-symmetric rank-$n$ Levi-Civita tensor
$\epsilon_{\mu_1\cdot\cdot\cdot\mu_n}$ is defined through
$\epsilon_{01\cdot\cdot\cdot(n-1)}=\sqrt{-g}$. Hence the Hodge duality
of $\star\textbf{F}$ is read off as
$\star\star\textbf{F}=(-1)^{pq+1}\textbf{F}$.
Apart from the
ordinary Hodge star operation, its various generalizations, as well
as their applications in physics, have been investigated in many works
(see \cite{ChungqdeH,MakikHodg,GenHoStors1,GenHoStors2,KriKMal,GuptaMal}, for example).
What is more, the usual exterior derivative of the $p$-form is presented by
${\rm d}\textbf{F}=(p!)^{-1}\partial_{\sigma}F_{\mu_1\cdot\cdot\cdot\mu_p}
{\rm d}x^{\sigma\mu_1\cdot\cdot\cdot\mu_p}$, directly leading to a significant
property of the exterior derivative that it gives zero when applied
twice in succession to an arbitrary differential form, i.e., it fulfills
the identity ${\rm d}^2=0$.

As what has been demonstrated in \cite{PengHg}, both the operators $O_1$ and $O_2$,
defined in terms of the Hodge star operation together with the exterior derivative and
expressed as
\be
O_1=\star {\rm d} \star {\rm d}\, , \quad \quad
O_2={\rm d}\star {\rm d}\star
\, , \label{O12}
\ee
respectively, can preserve the degree of the $p$-form field $\textbf{F}$. It is
easy to check that $O_1O_2=0=O_2O_1$ with the help of the identity
${\rm d}^2=0$. See \cite{PengHg,JFPleban,SharEg,SharEg2} for more information of the
properties about $O_1$ and $O_2$. Particularly, the works
\cite{JFPleban,SharEg,SharEg2} have explicitly exploited the properties of the
operator algebra generated by Hodge star and the exterior derivative. More
generally, through the linear combination of the operators $O_1^j$ and $O_2^k$,
we further propose an operator $\mathbb{O}_{jk}$ that is presented by
\footnote{We only focus on the combined operators depending on the Hodge star
operator and the exterior derivative in this paper, while Eq. (\ref{PoppForm})
shows that the degree-preserving d'Alembertian $\Box$ is also dependent of the
curvature tensors $R_{\mu\nu\rho\sigma}$ and $R_{\mu\nu}$ in curved
spacetimes. As a result, the d'Alembertian operator does not enter into
the definition of $\mathbb{O}_{jk}$.}
\be
\mathbb{O}_{jk}(\alpha_{1j},\alpha_{2k})=\alpha_{1j}O_1^j+\alpha_{2k}O_2^k
\, , \label{LinearComO12}
\ee
where both $\alpha_{1j}$ and $\alpha_{2k}$ are constant parameters (it is also allowed
that they are functions of spacetime coordinates). It is straightforward to
verify that $\mathbb{O}_{jk}$ still guarantees the degree of an arbitrary
$p$-form field to remain unchanged, namely,
$\mathbb{O}_{jk}:\Omega^p(M)\rightarrow \Omega^p(M)$, as well as
$\mathbb{O}^m_{jk}=\alpha_{1j}^mO_1^{mj}+\alpha_{2k}^mO_2^{mk}$. Making use
of the operator $\mathbb{O}^m_{jk}$ to act on $\textbf{F}$, we are
able to obtain a variety of $p$-form fields $\mathbb{O}^m_{jk}\textbf{F}$
that are the derivatives of order ${\rm Max}\{2mj,2mk\}$ of $\textbf{F}$.
What is more, without the consideration of the operations
$(\star\star)^k\in\{-1,1\}$, the combined operation
$\Sigma_{j,k}\mathbb{O}_{jk}$ could be viewed as
the most universal operator constructed from the combination of the Hodge
duality operation and the exterior derivative, which takes an
arbitrary $p$-form back into a certain $p$-form.

Subsequently, substituting $\alpha_{11}=(-1)^{np+1}$ and
$\alpha_{21}=(-1)^{n}\alpha_{11}$ into
the combined operator $\mathbb{O}_{11}(\alpha_{11},\alpha_{21})$, one can construct the
well-known Laplace-de Rham operator $\Delta$ (it is also called as Laplacian
or Laplace-de Rham operator in literature) from its action on the $p$-form field
$\textbf{F}$, written as \cite{TFrankel,EGHPhyR}
\bea
\Delta&=&(-1)^{np+1}O_1 + (-1)^{np+n+1}O_2 \nn \\
&=&\hat{\delta} {\rm d}+{\rm d}\hat{\delta}
\, . \label{ConO12}
\eea
Here the duality of the operation combining the exterior derivative and Hodge star
$\hat{\delta}=(-1)^{np+n+1}\star {\rm d} \star$
represents the well-known codifferential, or coderivative, or co-exterior derivative
\cite{TFrankel,GRStrauman,EGHPhyR},
which fulfills $\hat{\delta}^2=0$, $\hat{\delta}{\rm d}=(-1)^{np+1}O_1$
(or ${\rm d}\star{\rm d}=(-1)^{p}\star\hat{\delta}{\rm d}$) and
${\rm d}\hat{\delta}=(-1)^{np+n+1}O_2$, while we adopt $\hat{\delta}$ instead of the
conventional $\delta$ to denote the codifferential since we prefer to reserve the
latter for the variational symbol. The last equality of Eq. (\ref{ConO12})
apparently demonstrates that the Laplace-de Rham operator is just the
anticommutator of the codifferential and the exterior derivative, namely,
$\Delta=\{\hat{\delta},{\rm d}\}$. By making use of the equalities
$\hat{\delta}^2=0$ and ${\rm d}^2=0$, we obtain the important properties associated
with the three de Rham cohomological operators $({\rm d},\hat{\delta},\Delta)$
(they are essential ingredients involving in the famous Hodge decomposition
theorem \cite{TFrankel,EGHPhyR}),
including $\Delta^k=(\hat{\delta} {\rm d})^k+({\rm d}\hat{\delta})^k$, together with the
commutation relations $[\hat{\delta},{\rm d}]\neq0$, $[{\rm d},\Delta^k]=0$ and
$[\hat{\delta},\Delta^k]=0$. Moreover, the combination of the
coderivative with the Hodge star operation gives rise to
\bea
\star\hat{\delta}&=&(-1)^{p+1}{\rm d}\star \, , \quad
\star\hat{\delta}{\rm d}\star=(-1)^{pq+1}{\rm d}\hat{\delta}
\, , \nn \\
\hat{\delta}\star&=&(-1)^{p}\star{\rm d} \, , \quad
\star {\rm d}\hat{\delta}\star=(-1)^{pq+1}\hat{\delta}{\rm d}
\label{HodstCo}
\eea
according to their respective actions on arbitrary $p$-forms. More properties and
applications on the two operators $\Delta$ and $\hat{\delta}$
could be found in the works \cite{JFPleban,NavePhi,BiniChRu}.

By contrast with the exterior derivative, which increases the degree of a differential
form by one unit, the codifferential decreases that of a form by one. Specifically, its
operation on the $p$-form field $\textbf{F}$ sends this one to the $(p-1)$-form
\be
(\hat{\delta}\textbf{F})_{\mu_2\cdot\cdot\cdot\mu_p}
=\nabla^{\mu_1} F_{\mu_1\cdot\cdot\cdot\mu_p}
=({\rm div}_g\textbf{F})_{\mu_2\cdot\cdot\cdot\mu_p}
\, , \label{DefCoder}
\ee
that is to say, $\hat{\delta}$ is consistent with the usual divergence
operator ${\rm div}_g$. Furthermore, if the $p$-form $\textbf{F}$ is an exact form,
demanding that $\textbf{F}={\rm d}\textbf{A}$, where the field
$\textbf{A}\in\Omega^{p-1}(M)$ is a $(p-1)$-form, the coderivative of
$\textbf{F}$ becomes
\bea
\nabla^{\mu_1} F_{\mu_1\cdot\cdot\cdot\mu_p}
&=&(\hat{\delta}{\rm d}\textbf{A})_{\mu_2\cdot\cdot\cdot\mu_p} \nn \\
&=&(\Delta\textbf{A})_{\mu_2\cdot\cdot\cdot\mu_p}
-({\rm d}\hat{\delta}\textbf{A})_{\mu_2\cdot\cdot\cdot\mu_p}
\, , \label{DivF1}
\eea
in which, after some algebraic manipulations, the component expressions
for $\hat{\delta}{\rm d}\textbf{A}$ and ${\rm d}\hat{\delta}\textbf{A}$ are given by
\bea
(\hat{\delta}{\rm d}\textbf{A})_{\mu_2\cdot\cdot\cdot\mu_p}
&=&p\nabla^\sigma\nabla_{[\sigma} A_{\mu_2\cdot\cdot\cdot\mu_p]}\nn \\
&=& \Box A_{\mu_2\cdot\cdot\cdot\mu_p}-(p-1)\nabla^\sigma\nabla_{[\mu_2}
A_{|\sigma|\mu_3\cdot\cdot\cdot\mu_p]} \, , \nn \\
({\rm d}\hat{\delta}\textbf{A})_{\mu_2\cdot\cdot\cdot\mu_p}
&=&(p-1)\nabla_{[\mu_2}\nabla^\sigma A_{|\sigma|\mu_3\cdot\cdot\cdot\mu_p]}
\, , \label{DivF}
\eea
respectively. Here and henceforth, the covariant d'Alembertian operator with
respect to the metric tensor $g_{\mu\nu}$ is presented by
$\Box=g^{\mu\nu}\nabla_\mu\nabla_\nu$. As a convention, a pair of square
brackets on $p$ indices refer to anti-symmetrization over those indices with
the common factor of $(p!)^{-1}$, while the horizontal bars around an index
denote that this one remains out of the anti-symmetrization.

Let us make some discussions on the applications of Eq. (\ref{DivF}). In the
case where $\textbf{A}$ is an arbitrary scalar (0-form) $\phi(x)$, it yields
$\hat{\delta}{\rm d}\phi=\Box\phi$ and ${\rm d}\hat{\delta}\phi=0$. On the other hand,
for the case where $\textbf{A}$ is a vector field, Eq. (\ref{DivF}) gives
rise to the result
$(O_1\textbf{A})^{\mu}=2(-1)^{n}\nabla_\nu\nabla^{[\mu}A^{\nu]}$, ensuring
that the conserved current $J^\mu_K=2\nabla_\nu\nabla^{[\mu}\xi^{\nu]}$
involved in the well-known Komar integral \cite{Komar} can be alternatively
expressed as $\textbf{J}_K=-\hat{\delta} {\rm d}\boldsymbol{\xi}$ (its Hodge duality
gives rise to the usual $(n-1)$-form current). Here it is
worthwhile to note that it is unnecessary to restrict $\xi^\mu$ to a Killing
vector and it can be arbitrary, so $\textbf{J}_K$ covers the generalized
Komar current with respect to
an almost-Killing vector presented in \cite{JuFeng,RuPaBo} as a special case
(see the quite recent work \cite{FGWal} for some properties of the conserved
current associated with almost-Killing vectors).
One is able to check that the current naturally yields the divergence-free
equation $\hat{\delta}\textbf{J}_K=0$. Furthermore, we put forward a more general
conserved current associated with the arbitrary vector $\xi^\mu$, taking the form
\be
\textbf{J}_{(1)}=\sum_{i=1}\chi_i \big(\hat{\delta}{\rm d}\big)^i
\boldsymbol{\xi}
\, . \label{GenKomCurr}
\ee
Here and in what follows, $\chi_i$'s denote arbitrary constant parameters.
Hence the 2-form potential $\textbf{K}_{(2)}$ corresponding to
$\textbf{J}_{(1)}$ could be expressed as
$\textbf{K}_{(2)}=-\sum_{i=1}\chi_i  \big({\rm d}\hat{\delta}\big)^{i-1}
{\rm d}\boldsymbol{\xi}$
on basis of the relation $\textbf{J}_{(1)}=-\hat{\delta}\textbf{K}_{(2)}$.
In general, based upon an arbitrary $p$-form field $\textbf{F}$, the action
of the operator $O_1=(-1)^{np+1}\hat{\delta}{\rm d}$ on it renders the possibility
to construct an anti-symmetric $p$-index tensor
\be
\textbf{J}_{(p)}=\sum_{i=1}\chi_i O_1^i\textbf{F}
\, , \label{DivlesF}
\ee
which apparently obeys the constraint
of covariant divergencelessness $\hat{\delta}\textbf{J}_{(p)}=0$. Through a
replacement of $O_1$ with $O_2$ in $\textbf{J}_{(p)}$, one obtains a general closed $p$-form
$\textbf{J}_{(p)}(O_1\rightarrow O_2)$.

With the help of the operators $\Delta$ and $\hat{\delta}$, we are able to recast the
general operator $\mathbb{O}_{jk}$ given by Eq. (\ref{LinearComO12}) into
\be
\mathbb{O}_{jk}(\alpha_{1j},\alpha_{2k})=\alpha_{1j}\mathbb{P}^j-\alpha_{2k}\mathbb{P}^k
+\alpha_{2k}\Delta^k
\, , \label{OjkofTriaCo}
\ee
where $\mathbb{P}=\hat{\delta}{\rm d}=(-1)^{np+1}O_1$ for convenience, whose operation on a
differential form has been given by Eq. (\ref{DivF}). Within the above equation,
if both $\alpha_{1j}$ and $\alpha_{2k}$ are allowed to be the functions of the spacetime
coordinates, the properties $\mathbb{P}^j\phi =\Box^j\phi$ and
$\Delta^k\phi=\mathbb{P}^k\phi$ are useful to the computation of $\mathbb{O}_{jk}^m$.
For example, when $m=2$, we have
$\mathbb{O}_{jk}^2=\alpha_{1j}^2\mathbb{P}^{2j}+\alpha_{2k}^2(\Delta^{2k}
-\mathbb{P}^{2k})+\alpha_{1j}\Upsilon_{jk}$ with $\Upsilon_{jk}$ presented by
$\Upsilon_{jk}=(\Box^j\alpha_{1j})\mathbb{P}^{j}+(\Box^j\alpha_{2k})(\Delta^{k}
-\mathbb{P}^{k})$.
Moreover, Eq. (\ref{OjkofTriaCo}) implies that it is completely possible to
utilize the Laplace-de Rham operator $\Delta$ to manipulate differential
forms so as to assist with a new perspective upon $\mathbb{O}_{jk}$. To
demonstrate this, we address ourselves to such a problem. After letting
$\Delta$ operate on an arbitrary $p$-form $\textbf{F}$, we arrive at
the so-called Weitzenb\"{o}ck identity
\bea
\Delta\textbf{F}&=&\Box\textbf{F}+\mathbf{\Omega}(\textbf{F}) \, , \nn \\
\Box\textbf{F}&=&
\frac{1}{p!}\Box F_{\mu_1\cdot\cdot\cdot\mu_p}{\rm d}x^{\mu_1\cdot\cdot\cdot\mu_p}
\, , \label{PoppForm}
\eea
in which the $p$-form $\mathbf{\Omega}(\textbf{F})$, arising from the
property of the non-commutativity of the covariant derivative associated
with the (pseudo-)Riemannian geometry, takes the form
\be
\Omega_{\mu_1\cdot\cdot\cdot\mu_p}=-pR^\sigma_{[\mu_1}
F_{\mid\sigma\mid\mu_2\cdot\cdot\cdot\mu_p]}+\frac{p(p-1)}{2}
R^{\rho\sigma}_{[\mu_1\mu_2}
F_{|\rho\sigma|\mu_3\cdot\cdot\cdot\mu_p]}
\, . \label{DefOmega}
\ee
It should be pointed out that $\mathbf{\Omega}=0$ if $\textbf{F}$ is a
0-form (scalar). Throughout this work $R_{\rho\sigma\mu\nu}$ and
$R_{\rho\sigma}$ stand for the standard Riemann curvature tensor and
the Ricci tensor of metric respectively. Specifically, the former is defined through
$\big(\nabla_\rho\nabla_\sigma-\nabla_\sigma\nabla_\rho\big)V_\mu=R_{\rho\sigma\mu\nu}V^\nu$
for an arbitrary vector $V^\nu$ \cite{GRStrauman,Waldgr}. Eq. (\ref{PoppForm}) indicates that
the operation $\Delta$ could be generally expressed in terms of the d'Alembertian operator
and the curvature tensors. As a result, we obtain the commutation relations
\bea
\big[\hat{\delta}{\rm d},\Box\big]\textbf{F}&=&
\mathbf{\Omega}(\hat{\delta}{\rm d}\textbf{F})
-\hat{\delta}{\rm d}\mathbf{\Omega}(\textbf{F}) \, , \nn \\
\big[{\rm d}\hat{\delta},\Box\big]\textbf{F}&=&
\mathbf{\Omega}({\rm d}\hat{\delta}\textbf{F})
-{\rm d}\hat{\delta}\mathbf{\Omega}(\textbf{F})\, , \nn \\
\big[\Delta,\Box\big]\textbf{F}&=&\mathbf{\Omega}(\Box\textbf{F})
-\Box\mathbf{\Omega}(\textbf{F})
\, . \label{CRfordAled}
\eea
In particular, when the spacetime is Minkowskian, according to the
vanishing of the Riemann curvature tensor in such a spacetime, namely,
$R_{\rho\sigma\mu\nu}=0$, the operator $\Delta$ is identified with the standard wave
operator $\Box_\eta=\partial^\mu\partial_\mu$.

In the remainder of the present section, to provide a deeper understanding
about the operator $\Delta$, instead of the general
situations, we take into account of its applications in several concrete types
of fields.

First, when $\textbf{F}=f(x)\boldsymbol{\epsilon}$ is an arbitrary $n$-form,
we obtain $\Delta^k\textbf{F}=\boldsymbol{\epsilon}\Box^kf$ \cite{TFrankel}. In parallel,
the action of $\Delta^k$ on the scalar field $\phi(x)$ gives rise to
$\Delta^k \phi(x)=\Box^k\phi(x)$, from which one can get the massless
wave equation relevant to the scalar field $\Box\phi(x)=0$.

Second, if it is supposed that the $(p-1)$-form $\textbf{A}$ satisfies the
restriction $\mathbb{P}\textbf{A}=0$, we have the identity
$\Box\textbf{A}+\mathbf{\Omega}(\textbf{A})-{\rm d}\hat{\delta}\textbf{A}=0$.
Performing $\Delta$ on the closed $p$-form $\textbf{F}={\rm d}\textbf{A}$ further
yields
\be
\Delta \textbf{F}=(-1)^{np+n+1}O_2{\rm d}\textbf{A}=0
\, . \label{PdF}
\ee
This means that $\textbf{F}$ is a harmonic $p$-form. Hence Eq. (\ref{PoppForm})
gives the wave equation of the closed $p$-form
\be
\Box\textbf{F}=-\mathbf{\Omega}(\textbf{F})
\, . \label{WaveEqofcF}
\ee
For example, when $\textbf{F}$ is a closed and co-closed 2-form
$\textbf{F}_{(2)}=F_{\mu\nu}{\rm d}x^{\mu\nu}/2$, that is to say,
${\rm d}\textbf{F}_{(2)}=0$ together with $\hat{\delta}\textbf{F}_{(2)}=0$,
we obtain the wave equation for such
a field in the tensor form \cite{GRStrauman}
\be
\Box F_{\mu\nu}+R^\sigma_{\mu}F_{\nu\sigma}
-R^\sigma_{\nu}F_{\mu\sigma}+R^{\rho\sigma}_{\mu\nu}F_{\rho\sigma}=0
\, . \label{Triafor2F}
\ee
What is more, making use of Eq. (\ref{PoppForm}), we are able to reexpress
the well-known Proca equation
$\hat{\delta}{\rm d}\textbf{A}_{(1)}=m^2 \textbf{A}_{(1)}$,
which describes the co-closed vector field $\textbf{A}_{(1)}=A_\mu{\rm d}x^\mu$
with mass $m$, into the form
\be
\Box A_\mu-\big(R_{\mu}^\nu +m^2\delta^\nu_\mu\big)A_\nu=0
\, . \label{ProcaEq}
\ee
Specially, when the Ricci tensor $R_{\mu\nu}=\lambda g_{\mu\nu}$,
Eq. (\ref{ProcaEq}) becomes
$\Box\textbf{A}_{(1)}=(\lambda+m^2) \textbf{A}_{(1)}$.

Third, with the help of the first equation of Eq. (\ref{DivF}) and
Eq. (\ref{PoppForm}), one deduces the following commutation relationship
between the exterior derivative and the covariant d'Alembertian operator:
\be
\big[{\rm d},\Box\big]\textbf{A}=\mathbf{\Omega}({\rm d}\textbf{A})
+\frac{1}{(p-2)!}\nabla_{\mu_1}\nabla^\rho\nabla_{\mu_2}
A_{\rho\mu_3\cdot\cdot\cdot\mu_p}{\rm d}x^{\mu_1\cdot\cdot\cdot\mu_p}
\, . \label{RelddA}
\ee
For instance, when the $(p-1)$-form $\textbf{A}$ is a scalar $\phi$,
this equation leads to
\bea
\big[\nabla_\mu,\Box^m\big]\phi&=&
-R^\nu_\mu\sum_{k=0}^{m-1}\Box^k\nabla_\nu\Box^{m-k-1}\phi \nn \\
&&-\sum_{k=1}^{m-1}\big(\Box^kR^\nu_\mu\big)\nabla_\nu\Box^{m-k-1}\phi
\, , \label{CRfordBi}
\eea
where the arbitrary integer $m\geq 1$.
Apart from Eq. (\ref{RelddA}), the commutation relation between the
codifferential $\hat{\delta}$ and the d'Alembertian operator $\Box$ is
\be
\big[\hat{\delta},\Box\big]\textbf{F}=\mathbf{\Omega}(\hat{\delta}\textbf{F})
-\hat{\delta}\mathbf{\Omega}(\textbf{F})
\,  \label{HadeBoACR}
\ee
on basis of their action on an arbitrary $p$-form $\textbf{F}$.
Particularly, for an arbitrary vector $V^\mu$,
$\mathbf{\Omega}(\hat{\delta}\textbf{V})=0$. Thus we have
\be
\big[\hat{\delta},\Box^m\big]\textbf{V}=
-\sum^{m-1}_{k=0}\Box^k\hat{\delta}
\mathbf{\Omega}\big(\Box^{m-k-1}\textbf{V}\big)
\, . \label{CRCoBom}
\ee
Obviously, if the spacetime is Ricci-flat,
$\big[\hat{\delta},\Box^m\big]\textbf{V}=0$.
When $m=1$, the above equation becomes
\be
\big[\hat{\delta},\Box\big]\textbf{V}=\frac{1}{2}V^\mu\nabla_\mu R
+R_{\mu\nu}\nabla^{(\mu} V^{\nu)}
\, . \label{CoBoCRV}
\ee
If $V^\mu$ is a Killing vector $(\nabla^{(\mu} V^{\nu)}=0)$, we obtain
$V^\mu\nabla_\mu R=0$ since $2\Box\textbf{V}=\hat{\delta}{\rm d}\textbf{V}$
and $\hat{\delta}\textbf{V}=0$. That is to say, the Lie derivative of
the Ricci scalar along a Killing vector field disappears.

Forth, for another interesting case of acting $\Delta$ on an arbitrary Killing
vector $\xi^\mu$, by virtue of its null divergence
$\hat{\delta}\boldsymbol{\xi}=0$, we obtain
$({\rm d}\hat{\delta})^k\boldsymbol{\xi}=0$. This implies that
\be
\mathbb{P}^k\boldsymbol{\xi}=\Delta^k\boldsymbol{\xi}
\,  \label{PtriaXi}
\ee
always holds true for an arbitrary nonnegative integer $k$. As a result, application
of the first equation in Eq. (\ref{DivF}) and Eq. (\ref{PoppForm}) to the computation
of the above equation with $k=1$ leads to the property
$\Box\boldsymbol{\xi}=\mathbf{\Omega}(\boldsymbol{\xi})$
or $\Box\xi^\mu=-R^\mu_\sigma\xi^\sigma$ for Killing vectors.
According to such a property, we make use of the equation
$\big(\Delta\textbf{F}_{(1)}\big)^\mu=\Box F_{(1)}^\mu-R^\mu_\sigma F_{(1)}^\sigma$
repeatedly to obtain
\bea
(\Delta^1\boldsymbol{\xi})^\mu&=&-2\xi^{\sigma_1}R^\mu_{\sigma_1} \, ,\nn \\
(\Delta^2\boldsymbol{\xi})^\mu&=&-2\xi^{\sigma_2}\big(\Box R^\mu_{\sigma_2}
-2R^\mu_{\sigma_1}R^{\sigma_1}_{\sigma_2}\big)
\, , \nn \\
(\Delta^3\boldsymbol{\xi})^\mu&=&-2\xi^{\sigma_3}\big[\Box^2 R^\mu_{\sigma_3}
-3\Box \big(R^\mu_{\sigma_2}R^{\sigma_2}_{\sigma_3}\big)
+4R^\mu_{\sigma_1}R^{\sigma_1}_{\sigma_2}R^{\sigma_2}_{\sigma_3}\big]
\, . \label{PonKV}
\eea
On basis of Eq. (\ref{PonKV}), one can go on doing so to get $(\Delta^4\boldsymbol{\xi})$,
$(\Delta^5\boldsymbol{\xi})$, $\cdot\cdot\cdot$. As a matter of fact, $\Delta^k\boldsymbol{\xi}$
must be of the general form $(\Delta^k\boldsymbol{\xi})^\mu=-2\xi^\nu X^{\mu}_{(k)\nu}$,
where $X^{\mu}_{(k)\nu}$ consists of the terms made up of
$R^\rho_{\sigma}$, $\Box R^\rho_{\sigma}$, $\cdot\cdot\cdot$, $\Box^{k-1}R^\rho_{\sigma}$.
Therefore, if $\Delta^k\boldsymbol{\xi}=0$ holds for arbitrary Killing vectors, it is
demanded that $X^{\mu}_{(k)\nu}=0$, which enables one to construct equations containing
higher-order derivative terms of curvature tensors. For example, in the $k=2$ case, we
get the equation
$X^{\mu}_{(2)\nu}=\Box R^\mu_{\nu}-2R^\mu_{\sigma}R^{\sigma}_{\nu}=0$ \cite{PengHg}.
To the contrary, if the spacetime is Ricci-flat, namely, $R_{\mu\nu}=0$, we deduce
that $\Delta^k\boldsymbol{\xi}=0$ holds true for all Killing vectors. So we arrive at the
conclusion that any Killing vector in Ricci-flat spacetime is harmonic.

Finally, let us summarize the main novel results obtained in this section.
First, we give a generic degree-preserving combined operator $\mathbb{O}_{jk}$.
Second, we propose a generic conserved current $\textbf{J}_{(1)}$
asscociated with an arbitrary 1-form and a covariant divergence-free
p-form $\textbf{J}_{(p)}$. Third, we make use of differential forms to
derive the wave equation (\ref{WaveEqofcF}) for a harmonic $p$-form.
Fourth, we obtain the commutation relations given in Eqs. (\ref{CRfordAled}),
(\ref{CRfordBi}) and (\ref{CRCoBom}). Fifth, the action of the
operator $\Delta^k$ on an arbitrary Killing vector is explicitly analysed.

%%%%%%%%%%%%%%%%%%%%%%%%%%%%%%%%%%%%%%%%%%%%%%%%%%%%%%%%%%%%%%%%%%%%%%%%
\section{The construction of $n$-forms and their properties}\label{three}
%%%%%%%%%%%%%%%%%%%%%%%%%%%%%%%%%%%%%%%%%%%%%%%%%%%%%%%%%%%%%%%%%%%%%%%%

Within the present section, we shall utilize the operators
$\mathbb{O}_{jk}(\alpha_{1j},\alpha_{2k})$ in Eq. (\ref{LinearComO12}) to
construct $n$-forms $\textbf{L}_{\hat{m}\tilde{n}}(\textbf{F},\textbf{H})$
in terms of the $p$-form fields $\textbf{F}$ and $\textbf{H}$, as
well as their equivalents. To understand those, we are going to
investigate several special cases where the operators are specifically
$\Delta$ and $\mathbb{P}$. Table 2 in Appendix \ref{appendA} summarizes all
the $n$-forms.

Let us start with the action of the operators
\be
\hat{O}=\mathbb{O}_{jl}(\alpha_{1j},\alpha_{2l}) \, , \qquad
\tilde{O}=\mathbb{O}_{st}(\beta_{1s},\beta_{2t})
\, , \label{O11albe}
\ee
on the $p$-form fields $\textbf{F}$ and $\textbf{H}$. Due to the degree-preserving
property of the operator $\mathbb{O}_{jk}$, we are able to obtain $p$-form fields
$\hat{O}^i\textbf{F}$ ($i=0,1,\cdot\cdot\cdot,\hat{m}$) and
$\tilde{O}^k\textbf{H}$ ($k=0,1,\cdot\cdot\cdot,\tilde{n}$), where
$\hat{m}$ and $\tilde{n}$ represent arbitrary non-negative integers.
Consequently, the combination of these fields allows us to construct two new
$p$-form fields
$\hat{\textbf{F}}_{\hat{m}},\tilde{\textbf{H}}_{\tilde{n}}\in \Omega^p(M)$.
Both of them are read off as
\be
\hat{\textbf{F}}_{\hat{m}}=\sum^{\hat{m}}_{i=0}\gamma_i\hat{O}^i\textbf{F}
\, , \quad \quad
\tilde{\textbf{H}}_{\tilde{n}}
=\sum^{\tilde{n}}_{k=0}\lambda_k\tilde{O}^k\textbf{H}
\, , \label{hatFN}
\ee
respectively, where $\gamma_i$'s together with $\lambda_k$'s are coupling constants,
and it is set that both the operators $\hat{O}^0$ and $\tilde{O}^0$
are taken to be the identity operation $\textbf{1}$. Based upon the above two
$p$-form fields $\hat{\textbf{F}}_{\hat{m}}$ and $\tilde{\textbf{H}}_{\tilde{n}}$,
a general $n$-form $\textbf{L}_{\hat{m}\tilde{n}}(\textbf{F},\textbf{H})$ is further
defined through
\bea
\textbf{L}_{\hat{m}\tilde{n}}&=&
\hat{\textbf{F}}_{\hat{m}}\wedge \star\tilde{\textbf{H}}_{\tilde{n}}
=\sum^{\hat{m}}_{i=0}\sum^{\tilde{n}}_{k=0}\gamma_i\lambda_k\textbf{U}^{ik}
\, , \nn \\
\textbf{U}^{ik}&=&
\hat{O}^i\textbf{F}\wedge\star\tilde{O}^k\textbf{H}
\, . \label{nformLmn}
\eea
The above $\textbf{L}_{\hat{m}\tilde{n}}$ is one of our main desired results in
the present work. It is worth noting that the motivation to adopt
$\textbf{U}^{ik}$ in the construction of the
$n$-form $\textbf{L}_{\hat{m}\tilde{n}}(\textbf{F},\textbf{H})$ mainly
stems from the fact that $\textbf{U}^{ik}$ covers certain Lagrangians
associated with gauge fields. Apart from $\textbf{U}^{ik}$, one may wonder whether
$\hat{O}^i\textbf{F}\wedge\tilde{O}^k\textbf{H}$ can be adopted to construct
the $n$-form if $\textbf{H}$ is a $q$-form\footnote{We thank the anonymous
referee for putting forward this question.}. The answer is yes. This is
attributed to the fact that $\hat{O}^i\textbf{F}\wedge\tilde{O}^k\textbf{H}=
(-1)^{pq+1}\hat{O}^i\textbf{F}\wedge\star
\mathbb{O}_{ts}^k(\beta_{2t},\beta_{1s})\tilde{\textbf{H}}$, where the $p$-form
$\tilde{\textbf{H}}=\star\textbf{H}$ is the Hodge duality of $\textbf{H}$.
Obviously, $\hat{O}^i\textbf{F}\wedge\tilde{O}^k\textbf{H}$ can be recast into
the form
$(-1)^{pq+1}\hat{O}^i\textbf{F}\wedge\star\tilde{O}^k\tilde{\textbf{H}}$ if
the operator $\tilde{O}$ is redefined as
$\tilde{O}=\mathbb{O}_{ts}(\beta_{2t},\beta_{1s})$. To this point,
$\hat{O}^i\textbf{F}\wedge\tilde{O}^k\textbf{H}$ is equivalent to
$\textbf{U}^{ik}$. What is more, the $n$-form $\textbf{U}^{ik}$ can be alternatively
defined via $\textbf{U}^{ik}\rightarrow\tilde{\textbf{U}}^{ik}$, where
$\tilde{\textbf{U}}^{ik}=\tilde{O}^k\textbf{H}\wedge\star\hat{O}^i\textbf{F}$.

In accordance with $\textbf{L}_{\hat{m}\tilde{n}}=L_{\hat{m}\tilde{n}}\boldsymbol{\epsilon}$, one
obtains
\be
L_{\hat{m}\tilde{n}}=\sum^{\hat{m}}_{i=0}\sum^{\tilde{n}}_{k=0}\gamma_i\lambda_kU^{ik}
\, . \label{nformLmn2}
\ee
Here the scalar (or inner) product $U^{ik}$ between two differential
$p$-forms $\hat{O}^i\textbf{F}$ and $\tilde{O}^k\textbf{H}$ is defined through
the contraction of their components, namely,
\bea
U^{ik}&=&\big\langle\hat{O}^i\textbf{F}\cdot\tilde{O}^k\textbf{H}\big\rangle \nn \\
&=&\frac{1}{p!}\big(\hat{O}^i\textbf{F}\big)_{\mu_1\cdot\cdot\cdot\mu_p}
\big(\tilde{O}^k\textbf{H}\big)^{\mu_1\cdot\cdot\cdot\mu_p}
\, . \label{DefUik}
\eea
Obviously, $U^{ik}=U^{ki}$ if
$\hat{O}=\tilde{O}$ and $\textbf{H}=\textbf{F}$.

Next, for the purpose of simplicity, we attempt to construct an alternative
but equivalent formulation of the $n$-form
$\textbf{L}_{\hat{m}\tilde{n}}(\textbf{F},\textbf{H})$. As a warmup, we take
into consideration of an $n$-form ${\rm d}\textbf{B}\wedge\star\textbf{H}$.
Here and in what follows, $\textbf{B}$ denotes an arbitrary $(p-1)$-form
$\textbf{B}$. With the help of the codifferential, the $n$-form
${\rm d}\textbf{B}\wedge\star\textbf{H}$ can be expressed as
\be
{\rm d}\textbf{B}\wedge\star\textbf{H}=-\textbf{B}\wedge\star\hat{\delta}\textbf{H}
+{\rm d}(\textbf{B}\wedge\star\textbf{H})
\, , \label{BHrela1}
\ee
or equivalently
\be
\big\langle{\rm d}\textbf{B}\cdot\textbf{H}\big\rangle
=-\big\langle\textbf{B}\cdot\hat{\delta}\textbf{H}\big\rangle
+\frac{1}{(p-1)!}\nabla^{\mu_1}\big(H_{\mu_1\cdot\cdot\cdot\mu_p}
B^{\mu_2\cdot\cdot\cdot\mu_p}\big)
\, . \label{BHrela2}
\ee
From Eq. (\ref{BHrela1}), one observes that ${\rm d}\textbf{B}\wedge\star\textbf{H}$ differs
from $-\textbf{B}\wedge\star\hat{\delta}\textbf{H}$ only by an exact form or a total
derivative term. As a consequence of Eq. (\ref{BHrela2}), one gets
\bea
\hat{\delta}{\rm d}\textbf{F}\wedge\star\textbf{H}
&=&\textbf{F}\wedge\star\hat{\delta}{\rm d}\textbf{H}
+{\rm d}\boldsymbol{\Theta}_1\, , \nn \\
{\rm d}\hat{\delta}\textbf{F}\wedge\star\textbf{H}
&=&\textbf{F}\wedge\star{\rm d}\hat{\delta}\textbf{H}
+{\rm d}\boldsymbol{\Theta}_2
\, , \label{delddelFH}
\eea
with the boundary terms
$\boldsymbol{\Theta}_1,\boldsymbol{\Theta}_2\in \Omega^{n-1}(M)$
given by
\bea
\boldsymbol{\Theta}_1&=&
\textbf{H}\wedge\star{\rm d}\textbf{F}-\textbf{F}\wedge\star{\rm d}\textbf{H}
 \, , \nn \\
\boldsymbol{\Theta}_2&=&\hat{\delta}\textbf{F}\wedge\star\textbf{H}
-\hat{\delta}\textbf{H}\wedge\star\textbf{F}
\, , \label{ThetaFH12}
\eea
respectively. According to Eq. (\ref{delddelFH}), it is easy to check that both
the $n$-forms $\hat{\delta}{\rm d}\textbf{F}\wedge\star{\rm d}\hat{\delta}\textbf{H}$ and
${\rm d}\hat{\delta}\textbf{F}\wedge\star\hat{\delta}{\rm d}\textbf{H}$ are exact forms
by virtue of ${\rm d}\star\hat{\delta}=0$. Furthermore, computations based upon
Eq. (\ref{delddelFH}) give rise to \cite{EGHPhyR}
\bea
\Delta\textbf{F}\wedge\star \textbf{H}
&=&-{\rm d}\textbf{F}\wedge\star{\rm d}\textbf{H}
-\hat{\delta}\textbf{F}\wedge\star \hat{\delta}\textbf{H} \nn \\
&&+{\rm d}\big(\textbf{H}\wedge\star{\rm d}\textbf{F}
+\hat{\delta}\textbf{F}\wedge\star\textbf{H}\big) \nn \\
&=&\textbf{F}\wedge\star \Delta\textbf{H}
+{\rm d}(\boldsymbol{\Theta}_1+\boldsymbol{\Theta}_2)
\, , \label{TriaFH}
\eea
as well as
\be
\big\langle O^j_a\textbf{F}\cdot O^k_b\textbf{H}\big\rangle
=\delta_{ab}\big\langle\textbf{F}\cdot O^{j+k}_a\textbf{H}\big\rangle
+\nabla^\mu(\bullet)_\mu
\, , \label{O12LinnP}
\ee
where $a,b=1,2$ and $(\bullet)_\mu$ stands for the total derivative term.
Eq. (\ref{TriaFH}) implies that $\Delta\textbf{F}\wedge\star \textbf{H}$
must be an exact form provided that $\textbf{H}$ is a harmonic $p$-form.

Subsequently, with the help of Eq. (\ref{O12LinnP}), the $n$-form
$\textbf{L}_{\hat{m}\tilde{n}}(\textbf{F},\textbf{H})$
can be recast into the following equivalent form:
\be
\textbf{L}_{\hat{m}\tilde{n}}=\check{\textbf{L}}_{\check{m}\check{n}}
+{\rm d}\boldsymbol{\Theta}_{(n-1)}
\, , \label{Lmnequl}
\ee
in which the $(n-1)$-form $\boldsymbol{\Theta}_{(n-1)}$ represents some
boundary term while the $n$-form
$\check{\textbf{L}}_{\check{m}\check{n}}(\textbf{F},\textbf{H})$ is given by
\be
\check{\textbf{L}}_{\check{m}\check{n}}=
\sum^{\check{m}}_{i=0}\rho_i\check{\textbf{U}}_\Delta^{i}
+\sum^{\check{n}}_{k=0} \sigma_k\check{\textbf{U}}_{\mathbb{P}}^{k}
\,  \label{AnotherLmn}
\ee
with the $n$-forms $\check{\textbf{U}}_\Delta^{i}$ and
$\check{\textbf{U}}_{\mathbb{P}}^{k}$ defined through
\bea
\check{\textbf{U}}_\Delta^{i}&=&\textbf{F}\wedge\star \Delta^i\textbf{H}
\, , \nn \\
\check{\textbf{U}}_{\mathbb{P}}^{k}&=&
\textbf{F}\wedge\star \mathbb{P}^k\textbf{H}
\, , \label{cheUtriP}
\eea
respectively, where $\rho_i$'s and $\sigma_k$'s are constant parameters.
Eq. (\ref{Lmnequl}) shows that $\textbf{L}_{\hat{m}\tilde{n}}$ is determined
by $\check{\textbf{L}}_{\check{m}\check{n}}$ up to a surface term only. As
a consequence, ignoring the contribution from such a term, one can utilize
$\check{\textbf{L}}_{\check{m}\check{n}}$ as an alternative but equivalent
form of $\textbf{L}_{\hat{m}\tilde{n}}$. Apparently the former has the
great advantage of simplicity. What is more, when $\textbf{F}=\textbf{H}$,
under the transformation
$\textbf{F}\rightarrow \textbf{F}+{\rm d}\textbf{Y}$, where $\textbf{Y}$
is an arbitrary $(p-1)$-form, one observes that the $n$-forms
$\check{\textbf{U}}_\Delta^{i}(\textbf{F},\textbf{F})$
and $\check{\textbf{U}}_{\mathbb{P}}^{k}(\textbf{F},\textbf{F})$ transform as
\bea
\check{\textbf{U}}_\Delta^{i}
&\rightarrow&\check{\textbf{U}}_\Delta^{i}
+(2\textbf{F}+{\rm d}\textbf{Y})\wedge\star{\rm d}\mathbb{P}^i\textbf{Y}
+{\rm d}(\bullet) \, , \nn \\
\check{\textbf{U}}_{\mathbb{P}}^{k}
&\rightarrow& \check{\textbf{U}}_{\mathbb{P}}^{k}
+{\rm d}(\bullet)\, , \label{UDelPtraY}
\eea
respectively. If further provided that $\mathbb{P}\textbf{Y}=0$, Eq. (\ref{UDelPtraY})
leads to that $\check{\textbf{L}}_{\check{m}\check{n}}(\textbf{F},\textbf{F})$ behaves
like $\check{\textbf{L}}_{\check{m}\check{n}}\rightarrow
\check{\textbf{L}}_{\check{m}\check{n}}+{\rm d}(\bullet)$
under the aforementioned transformation.

Finally, in order to illustrate the structure of the $n$-form
$\textbf{L}_{\hat{m}\tilde{n}}(\textbf{F},\textbf{H})$, we move on to take into account three
typical cases where the operators $\hat{O}$ and $\tilde{O}$ take the
specific values $\Delta$ and $\mathbb{P}$. First,
let $\hat{O}=\Delta=\tilde{O}$. In such a case, $\textbf{U}^{ik}$ is denoted
by the notation $\textbf{U}_\Delta^{ik}$, taking the form
\bea
\textbf{U}_\Delta^{ik}&=&\Delta^i\textbf{F}\wedge\star\Delta^k\textbf{H} \nn \\
&=&\check{\textbf{U}}_\Delta^{i+k}
+{\rm d}\boldsymbol{\Theta}_\Delta^{ik}
\, , \label{UikTriang}
\eea
with the $n$-form $\check{\textbf{U}}_\Delta^{i+k}
=\textbf{F}\wedge\star \Delta^{i+k}\textbf{H}$ and
the surface term $\boldsymbol{\Theta}_\Delta^{ik}$ defined through
\be
\boldsymbol{\Theta}_\Delta^{ik}
=\sum_{j=1}^i\Big(\hat{\boldsymbol{\Theta}}_{\Delta}^{ik,j}
+\tilde{\boldsymbol{\Theta}}_{\Delta}^{ik,j}\Big)
\, , \label{Thetriang}
\ee
where the $(n-1)$-forms $\hat{\boldsymbol{\Theta}}_{\Delta}^{ik,j}$ and
$\tilde{\boldsymbol{\Theta}}_{\Delta}^{ik,j}$ $(1\leq j\leq i)$, derived
according to Eq. (\ref{delddelFH}) or (\ref{TriaFH}), are presented by
\bea
\hat{\boldsymbol{\Theta}}_{\Delta}^{ik,j}&=&
+\Delta^{k+j-1}\textbf{H}\wedge\star {\rm d}\Delta^{i-j}\textbf{F} \nn \\
&&-\Delta^{i-j}\textbf{F}\wedge\star {\rm d}\Delta^{k+j-1}\textbf{H}
 \, , \nn \\
\tilde{\boldsymbol{\Theta}}_{\Delta}^{ik,j}&=&
+\hat{\delta}\Delta^{i-j}\textbf{F}\wedge\star\Delta^{k+j-1}\textbf{H} \nn \\
&&-\hat{\delta}\Delta^{k+j-1}\textbf{H}\wedge\star\Delta^{i-j}\textbf{F}
\, , \label{ThetaTriag}
\eea
respectively. Neglecting surface terms, one straightforwardly
finds that $\textbf{L}_{\hat{m}\tilde{n}}$ is equivalently described by
\be
\check{\textbf{L}}^\Delta_{\check{m}}
=\sum^{\check{m}}_{i=0}\rho_i\check{\textbf{U}}_\Delta^{i}
\, . \label{Lmfortria}
\ee
Particularly, one obtains
$U_\Delta^{ik}=\langle\Box^i\textbf{F}\cdot\Box^k\textbf{H}\rangle$ when the spacetime
is Minkowskian. With imposition of the condition that $\textbf{F}={\rm d}\textbf{A}$, Eq. (\ref{PdF})
shows that performing $\Delta^i$ upon $\textbf{F}$ yields
$\Delta^i\textbf{F}=(-1)^{inp+in+i}O^i_2{\rm d}\textbf{A}$, demonstrating that the operator
$O_2$ acts on $\textbf{F}$ only. Therefore, due to Eq. (\ref{BHrela1}), the
$n$-form $\textbf{U}_\Delta^{ik}$ may be expressed as
\be
\textbf{U}_\Delta^{ik}=-\Delta^i\textbf{A}\wedge\star\mathbb{P}^{k}\hat{\delta}\textbf{H}
+{\rm d}\big(\Delta^i\textbf{A}\wedge\star\Delta^k\textbf{H}\big)
\, , \label{UikfordA}
\ee
or equivalently
\bea
U_\Delta^{ik}&=&-\big\langle\Delta^i\textbf{A}
\cdot\mathbb{P}^{k}\hat{\delta}\textbf{H}\big\rangle+\nabla^{\mu_1}B^{ik}_{\mu_1}
 \, , \nn \\
B^{ik}_{\mu_1}&=&\frac{1}{(p-1)!}(\Delta^k\textbf{H})_{\mu_1\cdot\cdot\cdot\mu_p}
(\Delta^i\textbf{A})^{\mu_2\cdot\cdot\cdot\mu_p}
\, . \label{UikfordA2}
\eea
For concreteness, considering as a simple example the $i,k=0$ and
$\textbf{H}=\textbf{F}$ case of Eq. (\ref{UikfordA}), we obtain
\bea
U_\Delta^{00}(\textbf{F},\textbf{F})
&=&\frac{1}{p!}F_{\mu_1\cdot\cdot\cdot\mu_p}
F^{\mu_1\cdot\cdot\cdot\mu_p} \nn \\
&=&-\big\langle\textbf{A}\cdot\mathbb{P}\textbf{A}\big\rangle
+\nabla^{\mu_1}B^{00}_{\mu_1}
\, , \label{U00forA}
\eea
which implies that our familiar $n$-form $\textbf{F}\wedge\star\textbf{F}$ can also
be expressed with respect to the operators $O_1$ and $O_2$. Second, in analogy to
the above-mentioned case, we take into account of replacing both the operators
$\hat{O}$ and $\tilde{O}$ in Eq. (\ref{nformLmn}) by the operator $\mathbb{P}$.
In this case, the $n$-form $\textbf{U}^{ik}$ in Eq. (\ref{nformLmn})
is of the form
\bea
\textbf{U}_\mathbb{P}^{ik}&=&
\mathbb{P}^i\textbf{F}\wedge\star\mathbb{P}^k\textbf{H} \nn \\
&=&\check{\textbf{U}}_{\mathbb{P}}^{i+k}
+{\rm d}\boldsymbol{\Theta}_\mathbb{P}^{ik} \, , \nn \\
\boldsymbol{\Theta}_\mathbb{P}^{ik}&=&
\sum_{j=1}^i\hat{\boldsymbol{\Theta}}_{\Delta}^{ik,j}
\big(\Delta\rightarrow \mathbb{P}\big)
\, , \label{UikP}
\eea
where the $n$-form $\check{\textbf{U}}_{\mathbb{P}}^{i+k}$ can be read off
from Eq. (\ref{cheUtriP}), that is, $\check{\textbf{U}}_{\mathbb{P}}^{i+k}
=\textbf{F}\wedge\star\mathbb{P}^{i+k}\textbf{H}$.
To obtain the surface term $\boldsymbol{\Theta}_\mathbb{P}^{ik}$,
Eq. (\ref{delddelFH}) has been used. Here note that
$\boldsymbol{\Theta}_\mathbb{P}^{ik}$ together with
$\boldsymbol{\Theta}_\Delta^{ik}$ could be used to express the surface term
$\boldsymbol{\Theta}_{(n-1)}$ given by Eq. (\ref{Lmnequl}).
Third, in the case where $\hat{O}=\Delta$ and
$\tilde{O}=\mathbb{P}$ (or $\hat{O}=\mathbb{P}$,
$\tilde{O}=\Delta$), we have
$\textbf{U}^{ik}=\textbf{U}_{\Delta,\mathbb{P}}^{ik}$
(or $\textbf{U}^{ik}=\textbf{U}_{\mathbb{P},\Delta}^{ik}$). Both of
them are written as
\bea
\textbf{U}_{\Delta,\mathbb{P}}^{ik}&=&
\Delta^i\textbf{F}\wedge\star\mathbb{P}^k\textbf{H}
\, , \nn \\
\textbf{U}_{\mathbb{P},\Delta}^{ik}&=&
\mathbb{P}^i\textbf{F}\wedge\star\Delta^k\textbf{H}
\, . \label{UikDeltaP}
\eea
In comparison with Eq. (\ref{UikP}), both the $n$-forms
$\textbf{U}_{\Delta,\mathbb{P}}^{ik}$  and
$\textbf{U}_{\mathbb{P},\Delta}^{ik}$ differ from
$\textbf{U}_\mathbb{P}^{ik}$ only by an exact form,
that is,
\bea
\textbf{U}_{\Delta,\mathbb{P}}^{ik}&=&\textbf{U}_\mathbb{P}^{ik}
+{\rm d}\big(\mathbb{P}^{i-1}\hat{\delta}\textbf{F}\wedge\star \mathbb{P}^{k}\textbf{H}\big)
 \, , \nn \\
\textbf{U}_{\mathbb{P},\Delta}^{ik}&=&\textbf{U}_\mathbb{P}^{ik}
+{\rm d}\big(\mathbb{P}^{k-1}\hat{\delta}\textbf{H}\wedge\star \mathbb{P}^{i}\textbf{F}\big)
\, . \label{UtpUp}
\eea
This means that $\textbf{U}_{\Delta,\mathbb{P}}^{ik}$ and
$\textbf{U}_{\mathbb{P},\Delta}^{ik}$ could be derived from
$\textbf{U}_\mathbb{P}^{ik}$, verifying the fact that the $n$-form
$\check{\textbf{L}}_{\check{m}\check{n}}(\textbf{F},\textbf{H})$
given by Eq. (\ref{AnotherLmn}) is only dependent on
$\check{\textbf{U}}_\Delta^{i}$ and
$\check{\textbf{U}}_{\mathbb{P}}^{k}$. For Eqs. (\ref{UikP})
and (\ref{UikDeltaP}), obviously, if it is assumed that
$\textbf{F}={\rm d}\textbf{A}$ and $\textbf{H}={\rm d}\textbf{B}$
hold, the $n$-forms
$\textbf{U}_\mathbb{P}^{ik}$, $\textbf{U}_{\Delta,\mathbb{P}}^{ik}$
and $\textbf{U}_{\mathbb{P},\Delta}^{ik}$ disappear. What is more,
comparing Eq. (\ref{UikfordA}) with Eq. (\ref{UikDeltaP}), we
establish the following expression:
\bea
\textbf{U}_{\Delta}^{ik}(\textbf{F},\textbf{H})
&=&-\textbf{U}_{\Delta,\mathbb{P}}^{i,k+1}(\textbf{A},\textbf{B})
+{\rm d}\big(\Delta^i\textbf{A}\wedge\star d\mathbb{P}^k \textbf{B}\big)
\nn \\
&=&-\textbf{U}_{\mathbb{P}}^{i,k+1}(\textbf{A},\textbf{B})
+{\rm d}\boldsymbol{\Theta}_{\textbf{A}\textbf{B}}^{ik}
\, , \label{RelatUtp}
\eea
with the surface term
$\boldsymbol{\Theta}_{\textbf{A}\textbf{B}}^{ik}
=\Delta^i\textbf{A}\wedge\star d\mathbb{P}^k \textbf{B}
-\mathbb{P}^{i-1}\hat{\delta}\textbf{A}
\wedge\star \mathbb{P}^{k+1}\textbf{B}$.
In light of Eqs. (\ref{UtpUp}) and (\ref{RelatUtp}), regardless of the
total derivative term, we come to the conclusion that
$\textbf{U}_{\Delta}^{ik}$, $\textbf{U}_\mathbb{P}^{i,k+1}$,
$\textbf{U}_{\mathbb{P},\Delta}^{i,k+1}$ and
$\textbf{U}_{\Delta,\mathbb{P}}^{i,k+1}$ are naturally equivalent with
each other.

With the $n$-form fields $\textbf{U}_\Delta^{ik}$, $\textbf{U}_\mathbb{P}^{ik}$,
$\textbf{U}_{\Delta,\mathbb{P}}^{ik}$ and $\textbf{U}_{\mathbb{P},\Delta}^{ik}$
in hand, actually, it is completely feasible to express the general $n$-form
$\textbf{U}^{ik}$ associated with the operators $\hat{O}$ and
$\tilde{O}$ through those fields. This is a direct consequence of the
fact that the operators $O^k_1=(-1)^{npk+k}\mathbb{P}^k$ and
$O^k_2= (-1)^{npk+nk+k}(\Delta^k-\mathbb{P}^k)$. Therefore, in order to get
the $n$-form $\textbf{L}_{\hat{m}\tilde{n}}$, one merely needs to carry out
computations for $\textbf{U}_\Delta^{ik}$,
$\textbf{U}_\mathbb{P}^{ik}$, $\textbf{U}_{\Delta,\mathbb{P}}^{ik}$ and
$\textbf{U}_{\mathbb{P},\Delta}^{ik}$ alternatively, while the latter two
could be derived from $\textbf{U}_\mathbb{P}^{ik}$ in virtue of
Eq. (\ref{UtpUp}). What is more, under the condition that
the contributions from the surface terms could be neglected, the equivalence
between $\textbf{L}_{\hat{m}\tilde{n}}$ and
$\check{\textbf{L}}_{\check{m}\check{n}}$ guarantees that
it is only necessary to evaluate $\check{\textbf{U}}_{\Delta}^{i}$ and
$\check{\textbf{U}}_{\mathbb{P}}^{k}$. If so, the calculations are
simplified notably.

%%%%%%%%%%%%%%%%%%%%%%%%%%%%%%%%%%%%%%%%%%%%%%%%%%%%%%%%%%%%%%%%%%%%%%%%%%%%%
\section{A potential application to the construction of an extended
Lagrangian associated to a $p$-form}
\label{four}
%%%%%%%%%%%%%%%%%%%%%%%%%%%%%%%%%%%%%%%%%%%%%%%%%%%%%%%%%%%%%%%%%%%%%%%%%%%%%

In this section, as an application of the aforementioned $n$-form
$\textbf{L}_{\hat{m}\tilde{n}}(\textbf{F},\textbf{H})$, we shall propose a
Lagrangian associated with an arbitrary $p$-form field $\textbf{A}_{(p)}$.
For simplicity, we will only focus on the detailed analysis about its equivalent
$\check{\textbf{L}}_{\check{m}\check{n}}(\textbf{F},\textbf{H})$ given by
Eq. (\ref{AnotherLmn}) instead of $\textbf{L}_{\hat{m}\tilde{n}}$. More
specifically, it is merely required to take into
consideration of the $n$-forms $\check{\textbf{U}}_{\Delta}^{i}$ and
$\check{\textbf{U}}_{\mathbb{P}}^{k}$. According to such a Lagrangian, we are
going to derive the equations of motion with respect to the fields.

As is well-known in the framework of gauge theory, for the Abelian
$(p-1)$-form gauge field $\textbf{A}$ with the field
strength $\textbf{F}={\rm d}\textbf{A}$, one popular form of its Lagrangian can
be presented by means of
\be
\textbf{L}_{00}=\gamma_0\lambda_0\textbf{F}\wedge \star\textbf{F}
\, , \label{maxwellLag}
\ee
which can be viewed as a generalization of the ordinary Lagrangian describing
Maxwell's theory of electromagnetism in four-dimensional Minkowski spacetime
and with the electromagnetic four-potential $A_\mu$. Further regardless of the
surface term in Eq. (\ref{U00forA}), which makes no contribution to the
equation of motion for the gauge field, one is able to reformulate the
Lagrangian (\ref{maxwellLag}) as
\bea
\textbf{L}_{00}&=&-\gamma_0\lambda_0\textbf{A}\wedge(\star\mathbb{P}\textbf{A})
\, , \nn \\
\textbf{L}_{00}&=&-\gamma_0\lambda_0\textbf{A}\wedge(\star\Delta\textbf{A})
\, . \label{LooinA}
\eea
In order to get the second equation in Eq. (\ref{LooinA}), we have imposed the
constraint that the field $\textbf{A}$ satisfies the Lorentz gauge condition
that $\textbf{A}$ is co-closed, namely,
$(\hat{\delta}\textbf{A})_{\mu_3\cdot\cdot\cdot\mu_p}
=\nabla^\rho A_{\rho\mu_3\cdot\cdot\cdot\mu_p}=0$,
or we apply Eq. (\ref{TriaFH}) to modify the Lagrangian (\ref{maxwellLag}) as
the one
\be
\tilde{\textbf{L}}_{00}=\gamma_0\lambda_0\big({\rm d}\textbf{A}\wedge
\star {\rm d}\textbf{A}
+\hat{\delta}\textbf{A}\wedge \star\hat{\delta}\textbf{A}\big)
\, , \label{ModFlagran}
\ee
which covers the gauge-fixed Lagrangians within the Minkowskian spacetime manifold
in \cite{GuptaMal}. Obviously, Eq. (\ref{LooinA}) demonstrates that the usual
Lagrangian for the $(p-1)$-form $\textbf{A}$ can be described by the $n$-form
$\textbf{L}_{\hat{m}\tilde{n}}(\textbf{A},\textbf{A})$ as well. In comparison with
Eq. (\ref{maxwellLag}), as we shall see later, it is of great convenience to derive
the equations of motion for the field in terms of the form of the Lagrangian given
by Eq. (\ref{LooinA}). What is more, for the well-known Lagrangian $L(\phi)$ with
respect to the scalar field $\phi$, usually given
by $L(\phi)=\sqrt{-g}\nabla^\mu\phi\nabla_\mu\phi$, it could be equivalently
of the form
\bea
\textbf{L}(\phi)&=&-\textbf{U}_{\mathbb{P}}^{01}(\phi,\phi) \nn \\
&=&-\phi(\star\mathbb{P}\phi)=-\phi(\star\Delta\phi)
\, \label{Triagphi}
\eea
without consideration of the contribution from the boundary term. That is to
say, in analogy to the Lagrangian of the field $\textbf{A}$, the one for the
scalar field might also be regarded as a special case of
$\textbf{L}_{\hat{m}\tilde{n}}$.

As a consequence, motivated by Eqs. (\ref{maxwellLag}), (\ref{LooinA}) and
(\ref{Triagphi}), here we propose that the $n$-form
$\textbf{L}_{\hat{m}\tilde{n}}(\textbf{F},\textbf{H})$ with
$\textbf{F}=\textbf{A}_{(p)}=\textbf{H}$ could be a generalized Lagrangian
associated with the $p$-form field $\textbf{A}_{(p)}$ from the mathematical
point of view. In comparison,
$\textbf{L}_{\hat{m}\tilde{n}}(\textbf{A}_{(p)},\textbf{A}_{(p)})$ makes
such extensions $\Delta\rightarrow\Delta^{i}$ and
$\mathbb{P}\rightarrow\mathbb{P}^{k}$ to the Lagranians (\ref{LooinA}) and
(\ref{Triagphi}). Since the boundary $\boldsymbol{\Theta}_{(p-1)}$ in
Eq. (\ref{Lmnequl}) makes no contribution to the equations of motion for the
fields, it is completely advisable for us to adopt the more convenient $n$-form
$\check{\textbf{L}}_{\check{m}\check{n}}\big(\textbf{A}_{(p)},\textbf{A}_{(p)}\big)$
presented by Eq. (\ref{AnotherLmn}) rather than
$\textbf{L}_{\hat{m}\tilde{n}}(\textbf{A}_{(p)},\textbf{A}_{(p)})$ as
the form of the Lagrangian.

In the remainder of this section, on basis of the Lagrangian
$\check{\textbf{L}}_{\check{m}\check{n}}\big(\textbf{A}_{(p)},\textbf{A}_{(p)}\big)$,
we take into account of the derivation for the Euler-Lagrange equation of motion
associated with the $p$-form $\textbf{A}_{(p)}$. To do this, on one hand, via
varying the $n$-form
$\check{\textbf{U}}_{\Delta}^{i}\big(\textbf{A}_{(p)},\textbf{A}_{(p)}\big)$
with respect to $\textbf{A}_{(p)}$, we obtain
\be
\delta\check{\textbf{U}}_\Delta^{i}=2\delta\textbf{A}_{(p)}\wedge\star
\Delta^i\textbf{A}_{(p)}
+\sum_{j=1}^i d\boldsymbol{\Psi}_\Delta^{ij}
\, , \label{VarcheUtri}
\ee
where the surface terms $\boldsymbol{\Psi}_\Delta^{ij}$ are given by
\bea
\boldsymbol{\Psi}_\Delta^{ij}&=&
+\Delta^{j-1}\textbf{A}_{(p)}\wedge\star {\rm d}\Delta^{i-j}\delta\textbf{A}_{(p)} \nn \\
&&-\Delta^{i-j}\delta\textbf{A}_{(p)}\wedge\star {\rm d}\Delta^{j-1}\textbf{A}_{(p)} \nn \\
&&+\hat{\delta}\Delta^{i-j}\delta\textbf{A}_{(p)}\wedge\star\Delta^{j-1}\textbf{A}_{(p)} \nn \\
&&-\hat{\delta}\Delta^{j-1}\textbf{A}_{(p)}\wedge\star\Delta^{i-j}\delta\textbf{A}_{(p)}
\, . \label{PsiTriaij}
\eea
If $\delta\check{\textbf{U}}_\Delta^{i}=0$, we get the equations of motion
$\Delta^i\textbf{A}_{(p)}=0$, demonstrating that a simple solution is the
closed and co-closed form $\textbf{A}_{(p)}$ satisfying
${\rm d}\textbf{A}_{(p)}=0=\hat{\delta}\textbf{A}_{(p)}$. For instance, when
$i=1$, we have the field equation for the 2-form $\textbf{A}_{(2)}$ with
the help of Eq. (\ref{PoppForm}), that is,
\be
\Box A_{\mu\nu}=R^\sigma_{\mu}A_{\sigma\nu}
-R^\sigma_{\nu}A_{\sigma\mu}-R^{\rho\sigma}_{\mu\nu}A_{\rho\sigma}
\, , \label{EMfor1A2}
\ee
which could be interpreted as the field equation associated with the Kalb-Ramond
Lagrangian $\textbf{L}_{KR}={\rm d}\textbf{A}_{(2)}\wedge\star{\rm d}\textbf{A}_{(2)}/2$
\cite{KalbRam} under the divergence-free gauge condition $\hat{\delta}\textbf{A}_{(2)}=0$,
and when $i=2$, the equation of motion for the 1-form $\textbf{A}_{(1)}$ is read off as
\be
\Box^2 A_{\mu}=2R^\sigma_{\mu}\Box A_{\sigma}+A_{\sigma}\Box R^\sigma_{\mu}
-R^\rho_{\mu}R_{\rho}^{\sigma}A_{\sigma}
\, . \label{EMfor2A1}
\ee
On the other hand, in an analogous way to the above-mentioned analysis on
$\delta\check{\textbf{U}}_\Delta^{i}$, let us deal with the variation of the $n$-form
$\check{\textbf{U}}_{\mathbb{P}}^{k}\big(\textbf{A}_{(p)},\textbf{A}_{(p)}\big)$
with respect to the field $\textbf{A}_{(p)}$. This gives rise to
\be
\delta\check{\textbf{U}}_{\mathbb{P}}^{k}=
2\delta\textbf{A}_{(p)}\wedge\star \mathbb{P}^k\textbf{A}_{(p)}
+\sum_{j=1}^k {\rm d}\boldsymbol{\Psi}_{\mathbb{P}}^{kj}
\, , \label{VarcheUP}
\ee
with the boundary terms $\boldsymbol{\Psi}_{\mathbb{P}}^{kj}$ defined through
\bea
\boldsymbol{\Psi}_{\mathbb{P}}^{kj}&=&
+\mathbb{P}^{j-1}\textbf{A}_{(p)}\wedge\star {\rm d}\mathbb{P}^{k-j}\delta\textbf{A}_{(p)}\nn \\
&&-\mathbb{P}^{k-j}\delta\textbf{A}_{(p)}\wedge\star{\rm d}\mathbb{P}^{j-1}\textbf{A}_{(p)}
\, . \label{PsiPkj}
\eea
Similarly, $\delta\check{\textbf{U}}_{\mathbb{P}}^{k}=0$ yields the equations of motion
$\mathbb{P}^{k}\textbf{A}_{(p)}=0$, according to which one gains
$\mathbb{P}^{k-1}\hat{\delta}\textbf{F}_{(p+1)}=0$ and $\Delta^k\textbf{F}_{(p+1)}=0$,
where the closed $(p+1)$-form $\textbf{F}_{(p+1)}={\rm d}\textbf{A}_{(p)}$. Letting us specialize
to the case where $k=1$ and $\textbf{A}_{(p)}$ is the aforementioned $(p-1)$-form
$\textbf{A}$, we observe that
\be
(\mathbb{P}\textbf{A})_{\mu_2\cdot\cdot\cdot\mu_p}=\nabla^{\mu_1}
F_{\mu_1\cdot\cdot\cdot\mu_p}=0
\, , \label{EMofAp}
\ee
which is just the field equation relevant for the Lagrangian (\ref{maxwellLag}).
Here $\textbf{F}={\rm d}\textbf{A}$ as before. In the assumption that the Lorentz
gauge condition $\hat{\delta}\textbf{A}=0$ holds, Eq. (\ref{EMofAp})
further transforms to $\Delta\textbf{A}=0$ or
$\Box A_\mu=R^\sigma_{\mu}A_{\sigma}$.

According to the fact that the Lagrangian
$\check{\textbf{L}}_{\check{m}\check{n}}\big(\textbf{A}_{(p)},\textbf{A}_{(p)}\big)$
is the linear combination of the $n$-forms
$\check{\textbf{U}}_{\Delta}^{i}\big(\textbf{A}_{(p)},\textbf{A}_{(p)}\big)$
and $\check{\textbf{U}}_{\mathbb{P}}^{k}\big(\textbf{A}_{(p)},\textbf{A}_{(p)}\big)$,
we find that the variations of $\check{\textbf{U}}_{\Delta}^{i}$
and $\check{\textbf{U}}_{\mathbb{P}}^{k}$
are sufficient for the derivation of the field equations related to the Lagrangian
$\check{\textbf{L}}_{\check{m}\check{n}}\big(\textbf{A}_{(p)},\textbf{A}_{(p)}\big)$.
As a result, we make use of Eqs. (\ref{VarcheUtri}) and (\ref{VarcheUP}) to
vary
$\check{\textbf{L}}_{\check{m}\check{n}}\big(\textbf{A}_{(p)},\textbf{A}_{(p)}\big)$
with respect to the field $\textbf{A}_{(p)}$ and obtain the equation of motion
\be
\sum^{\check{m}}_{i=0}\rho_i\Delta^i\textbf{A}_{(p)}
+\sum^{\check{n}}_{k=0} \sigma_k\mathbb{P}^{k}\textbf{A}_{(p)}=0
\, . \label{EMfromcheL}
\ee
Apparently, the left hand side of the above equation results from the linear combination of
the $p$-forms $\Delta^i\textbf{A}_{(p)}$ and $\mathbb{P}^{k}\textbf{A}_{(p)}$.

Ultimately, we have to point out that it is of great importance to verify the 
stability with respect to the Lagrangian
$\check{\textbf{L}}_{\check{m}\check{n}}\big(\textbf{A}_{(p)},\textbf{A}_{(p)}\big)$.
In fact, since this Lagrangian includes terms with higher-order time derivatives, it
generally encounters the ghost-like instability referred to as the Ostrogradsky
instability\footnote{We thank the anonymous referee for pointing out this.},
which is a linear instability existing in the Hamiltonian associated with a
non-degenerate Lagrangian containing time derivative terms higher than the
first order \cite{Oinstab}. As is known, for the purpose of constructing
well-behaved scalar-tensor theories involved in higher-order derivatives,
several approaches have been proposed to avoid the Ostrogradsky instabilities
of these theories (see works \cite{OinGali,OinHorde,LanNo},
for instance). Such methods maybe provide some insight into the avoidance of
the linear instability of the Hamiltonian related to the Lagrangian
$\check{\textbf{L}}_{\check{m}\check{n}}$, which is left for future work.

%%%%%%%%%%%%%%%%%%%%%%%%%%%%%%
\section{Conclusions}\label{five}
%%%%%%%%%%%%%%%%%%%%%%%%%%%%%%

In the present paper, we have systematically investigated the properties of the operators
that are able to generate differential forms maintaining the invariance of the degree for
an arbitrary $p$-form field, as well as their applications in constructions of $n$-forms
and Lagrangians associated with $p$-form fields. More explicitly, through the linear
combination of the operators $O_1$ and $O_2$, which are made up of the Hodge star
operation and the exterior derivative, we have obtained the general operator
$\mathbb{O}_{jk}(\alpha_{1j},\alpha_{2k})$ given by Eq. (\ref{LinearComO12})
or (\ref{OjkofTriaCo}). With the help of $O_1$, a new conserved $p$-form
$\textbf{J}_{(p)}$ is presented in Eq. (\ref{DivlesF}). In order to understand
the operators preserving the form degree, we have given detailed analysis
on the Laplace-de Rham operator $\Delta$, as well as commutation relations
between two operations and the action of $\Delta^k$ on
an arbitrary Killing vector. Particularly, the concrete relationship between
the Laplace-de Rham operator and the d'Alembertian operator
has been established through Eq. (\ref{PoppForm}). Subsequently,
in terms of the actions of the operator $\mathbb{O}_{jk}$ on the two
$p$-forms $\textbf{F}$ and $\textbf{H}$, the general $p$-forms
$\hat{\textbf{F}}_{\hat{m}}$ and $\tilde{\textbf{H}}_{\tilde{n}}$ have been presented
via Eq. (\ref{hatFN}). Based upon them, the general $n$-form
$\textbf{L}_{\hat{m}\tilde{n}}(\textbf{F},\textbf{H})$ in Eq. (\ref{nformLmn}) has further
been constructed, and its three special cases where the operators are specifically the ones
$\Delta$ and $\mathbb{P}$ have been discussed in detail. As a matter of fact, we have
demonstrated that $\textbf{L}_{\hat{m}\tilde{n}}$ can be expressed as an alternative
but equivalent $n$-form $\check{\textbf{L}}_{\check{m}\check{n}}(\textbf{F},\textbf{H})$
given by Eq. (\ref{AnotherLmn}) without the contribution from the total derivative term.

Finally, inspired by the forms (\ref{LooinA}) and (\ref{Triagphi}) of the usual
Lagrangians for the $p$-forms and scalar fields, we suggest that the $n$-form
$\textbf{L}_{\hat{m}\tilde{n}}\big(\textbf{A}_{(p)},\textbf{A}_{(p)}\big)$
or $\check{\textbf{L}}_{\check{m}\check{n}}\big(\textbf{A}_{(p)},\textbf{A}_{(p)}\big)$,
including the higher-order derivative $p$-forms
$\Delta^{i}\textbf{A}_{(p)}$ and $\mathbb{P}^{k}\textbf{A}_{(p)}$, could be thought
of as a higher-order derivative generalization of the usual Lagrangian related
to the $p$-form $\textbf{A}_{(p)}$ in the mathematical point of view. In terms of
the Lagrangian, we have derived the equations of motion for the fields by
varying the $n$-forms
$\check{\textbf{U}}_{\Delta}^{i}\big(\textbf{A}_{(p)},\textbf{A}_{(p)}\big)$,
$\check{\textbf{U}}_{\mathbb{P}}^{k}\big(\textbf{A}_{(p)},\textbf{A}_{(p)}\big)$
and $\check{\textbf{L}}_{\check{m}\check{n}}\big(\textbf{A}_{(p)},\textbf{A}_{(p)}\big)$
with respect to $\textbf{A}_{(p)}$. However, apart from the mathematical
aspects of the extended Lagrangian
$\check{\textbf{L}}_{\check{m}\check{n}}\big(\textbf{A}_{(p)},\textbf{A}_{(p)}\big)$,
it is of great importance to seek the physical understandings behind the Lagrangian,
for instance, the Ostrogradsky instability arising from the higher-order 
time derivatives in the Lagrangian. This remains to be investigated in the future research. Besides,
the potential applications of the conserved $p$-form $\textbf{J}_{(p)}$ in
theories involving higher derivatives of fields are deserved to be
investigated in future.

\section*{Acknowledgments}

We would like to thank the anonymous referees for their valuable suggestions
and comments. This work was supported by the Natural Science Foundation of
China under Grant Nos. 11865006 and 11505036. It was also partially supported
by the Technology Department of Guizhou province Fund under Grant
Nos. [2018]5769 and [2016]1104.

%%%%%%%%%%%%%%%%%%%%%%%%%%%%%%%%%%%%%%%%%%%%%%%%%%%%%%%%%%%
\appendix
%%%%%%%%%%%%%%%%%%%%%%%%%%%%%%%%%%%%%%%%%%%%%%%%%%%%%%%%
\section{Notations and conventions} \label{appendA}
%%%%%%%%%%%%%%%%%%%%%%%%%%%%%%%%%%%%%%%%%%%%%%%%%%%%

Throughout this paper, the positive integer $n$ represents the dimension of
the spacetime. The non-negative integers $p$ and $q=n-p$ stand for the form
degree. As usual, the tensor (or spacetime coordinate) indices will be
labeled by the Greek letters $\mu$, $\nu$, $\mu_1$, $\nu_1$, $\mu_2$,
$\nu_2$, $\cdot\cdot\cdot$, $\rho$, $\sigma$. Each of them runs from 0
to $(n-1)$. All the Latin indices
$i$, $j$, $k$, $l$, $s$, $t$, $m$, $\check{m}$ and $\check{n}$ are
non-negative integers, and they will be used to represent exponents of
the operators and labels. The quantities $\alpha_{1j}$, $\alpha_{2k}$,
$\beta_{1s}$, $\beta_{2t}$, $\gamma_i$, $\lambda_k$, $\chi_i$, $\rho_i$ and
$\sigma_k$ denote arbitrary constant parameters. Like in \cite{Waldgr},
we use boldface letters to denote differential forms to avoid confusion with
functions.

Table \ref{opetab} displays the definitions for the main operators appearing in
this paper.
\begin{table}[H]
\caption{Directory of operators}
    \vspace{15pt}
    \centering
    \begin{tabular}{p{1.55cm}p{5.00cm}}
        \hline
        \hline
     Operator  & Definition     \\
        \hline
        $\star$        & Hodge star, given by Eq. (\ref{DefHodge})       \\
        ${\rm d}$      & exterior derivative               \\
        $\hat{\delta}$ & codifferential:
                         $(-1)^{np+n+1}\star{\rm d}\star$   \\
        $\Delta$  &Laplace-de Rham: $\hat{\delta} {\rm d}+{\rm d}\hat{\delta}$    \\
        $\Box$       & d'Alembertian: $g^{\mu\nu}\nabla_\mu\nabla_\nu$ \\
        $\mathbb{P}$  & $\hat{\delta}{\rm d}$                \\
        $O_1$  & $\star {\rm d} \star {\rm d}$                \\
        $O_2$  & ${\rm d}\star {\rm d}\star$              \\
        $\mathbb{O}_{jk}$  & $\alpha_{1j}O_1^j+\alpha_{2k}O_2^k$ \\
        $\hat{O}$  & $\alpha_{1j}O_1^j+\alpha_{2l}O_2^l$    \\
        $\tilde{O}$  &$\beta_{1s}O_1^s+\beta_{2t}O_2^t$ \\

      \hline
    \end{tabular}
    \label{opetab}
\end{table}

Through the action of the degree-preserving operators on the $p$-forms
$\textbf{F}$ and $\textbf{H}$, we have obtained some novel $n$-forms,
which are displayed by Table \ref{nforms}.
\begin{table}[H]
\caption{Definitions for $n$-forms}
    \vspace{15pt}
    \centering
    \begin{tabular}{p{1.2cm}p{4.2cm}}
        \hline
        \hline
     $n$-form  & Expression     \\
        \hline
        $\textbf{U}^{ik}$
        &$\hat{O}^i\textbf{F}\wedge\star\tilde{O}^k\textbf{H}$ \\
        $\check{\textbf{U}}_\Delta^{i}$
        &$\textbf{F}\wedge\star \Delta^i\textbf{H}$ \\
        $\check{\textbf{U}}_{\mathbb{P}}^{k}$
        &$\textbf{F}\wedge\star \mathbb{P}^k\textbf{H}$    \\
        $\textbf{U}_\Delta^{ik}$
        &$\Delta^i\textbf{F}\wedge\star\Delta^k\textbf{H}$ \\
        $\textbf{U}_\mathbb{P}^{ik}$
        &$\mathbb{P}^i\textbf{F}\wedge\star\mathbb{P}^k\textbf{H}$\\
        $\textbf{U}_{\Delta,\mathbb{P}}^{ik}$
        &$\Delta^i\textbf{F}\wedge\star\mathbb{P}^k\textbf{H}$ \\
        $\textbf{U}_{\mathbb{P},\Delta}^{ik}$
        &$\mathbb{P}^i\textbf{F}\wedge\star\Delta^k\textbf{H}$\\
        $\textbf{L}_{\hat{m}\tilde{n}}$
        &$\sum^{\tilde{n}}_{k=0}\gamma_i\lambda_k\textbf{U}^{ik}$    \\
        $\check{\textbf{L}}_{\check{m}\check{n}}$
        &$\sum^{\check{m}}_{i=0}\rho_i\check{\textbf{U}}_\Delta^{i}
         +\sum^{\check{n}}_{k=0} \sigma_k\check{\textbf{U}}_{\mathbb{P}}^{k}$\\

      \hline
    \end{tabular}
    \label{nforms}
\end{table}

\end{document}